\documentclass[preprint,12pt, a4paper]{elsarticle}

\usepackage{amssymb}
\usepackage{hyperref}
\usepackage{enumitem}
\usepackage{verbatim}
\usepackage{fancyvrb}
\usepackage{listings}

\journal{v1.0}

\begin{document}
\renewcommand{\labelenumii}{\arabic{enumi}.\arabic{enumii}}

\begin{frontmatter}



\title{SUS-Lib: An automated tool for usability evaluation based on the Software Usability Scale from user feedback}


\author[label1]{Paweł Weichbroth}
\author[label2]{Małgorzata Giedrowicz}
\address[label1]{Gdansk University of Technology, Faculty of Electronics, Telecommunications and Informatics, Department of Software Engineering, Narutowicza 11/12, pawel.weichbroth@pg.edu.pl}
\address[label2]{Meritus Systemy Informatyczne, Prosta 70, Warsaw, Poland, malgorzatag@meritus.pl}

\begin{abstract}
Usability evaluation has received considerable attention from both the research and practice communities. While there are many evaluation tools available, the Software Usability Scale (SUS) is the most widely used. In this paper, we introduce and describe the SUS-Lib software package, which aims to compute SUS scores and generate graphical figures based on user input. SUS-Lib responds to the need for user-friendly software that requires only basic knowledge and skills of the Python environment and command line tools. By using open source solutions and low hardware resources, SUS-Lib is a cost-effective solution. In addition, due to its generic nature, SUS-Lib can also be used in different research setups and settings.
\end{abstract}


\end{frontmatter}


\section*{Metadata}
\label{sec:metadata}

\begin{table}[!h]
\begin{tabular}{|l|p{6.5cm}|p{6.5cm}|}
\hline
C1 & Current code version & v1 \\
\hline
C2 & Permanent link to code/repository used for this code version & \url{https://github.com/MalgiGie/SUS-library} \\
\hline
C3  & Permanent link to Reproducible Capsule & -- \\
\hline
C4 & Legal Code License   & MIT License\\
\hline
C5 & Code versioning system used & git\\
\hline
C6 & Software code languages, tools, and services used & Python\\
\hline
C7 & Compilation requirements, operating environments \& dependencies & Python 3.6 or higher, matplotlib, numpy\\
\hline
C8 & If available Link to developer documentation/manual &  -- \\
\hline
C9 & Support email for questions & malgigie@gmail.com\\
\hline
\end{tabular}
\caption{Code metadata}
\label{codeMetadata} 
\end{table}

\begin{table}[!h]
\begin{tabular}{|l|p{6.5cm}|p{6.5cm}|}
\hline
S1 & Current software version & 1.1 \\
\hline
S2 & Permanent link to executables of this version  & -- \\
\hline
S3  & Permanent link to Reproducible Capsule & -- \\
\hline
S4 & Legal Software License & -- \\
\hline
S5 & Computing platforms/Operating Systems & Microsoft Windows\\
\hline
S6 & Installation requirements \& dependencies & --\\
\hline
S7 & If available, link to user manual - if formally published include a reference to the publication in the reference list & \url{https://github.com/MalgiGie/SUS-library/blob/main/README.md} \\
\hline
S8 & Support email for questions & malgigie@gmail.com\\
\hline
\end{tabular}
\caption{Software metadata}
\label{executabelMetadata} 
\end{table}

\section{Motivation and significance}
Generally speaking, in the digital product landscape, usability is of vital importance and has a direct impact on user satisfaction and adoption. While high usability means that users can achieve their goals without confusion, frustration, or unnecessary effort, on the other hand, poor usability can lead to frustration, errors, and inefficiencies that drive users away. In this sense, measuring usability is necessary at every stage of the software lifecycle. 

According to ISO 9241-11, by definition, usability is the "extent to which a system, product, or service can be used by specified users to achieve specified goals with effectiveness, efficiency, and satisfaction in a specified context of use" \cite{ISO9242-11}.
Over the years, there have been many attempts to develop effective tools for usability evaluation. Among them, the most recognized, used, and adopted tool is the Software Usability Scale (SUS).

Developed by John Brooke in 1986 while working at Digital Equipment Corporation \cite{brooke1996sus}, the SUS is a simple, questionnaire-based method that allows users to provide quick feedback on various aspects of a system. The results are easy to interpret and provide a standardized measure of overall usability. 

SUS consists of ten following items (statements).

\begin{enumerate}
    \item[($i_1$):] I think that I would like to use this system frequently.
    \item[($i_2$):] I found the system unnecessarily complex.
    \item[($i_3$):] I thought the system was easy to use.
    \item[($i_4$):] I think that I would need the support of a technical person to be able to use this system.
    \item[($i_5$):] I found that the various functions in this system were well integrated.
    \item[($i_6$):] I thought there was too much inconsistency in this system.
    \item[($i_7$):] I imagine that most people would learn to use this system very quickly.
    \item[($i_8$):] I found the system very cumbersome to use.
    \item[($i_9$):] I felt very confident using the system.
    \item[($i_{10}$):] I needed to learn a lot of things before I could get going with this system.
\end{enumerate}

A user responds by choosing from a predefined five-point Likert scale. The scale ranges from strongly disagree (1) to strongly agree (5). The SUS score is calculated in the following way:
\begin{itemize}
    \item each item’s score contribution ranges from 0 to 4;
    \item for the positive items (odd numbers: 1, 3, 5, 7, and 9), the score contribution is the scale position minus 1;
    \item for the negative items (even numbers: 2, 4, 6, 8, and 10), the score contribution is 5 minus the scale position;
    \item these two new scores are added up the total score, and then multiplied this by 2.5 to get the SUS value.
\end{itemize}

\begin{figure}[h]
\centering
\includegraphics[width=12cm]{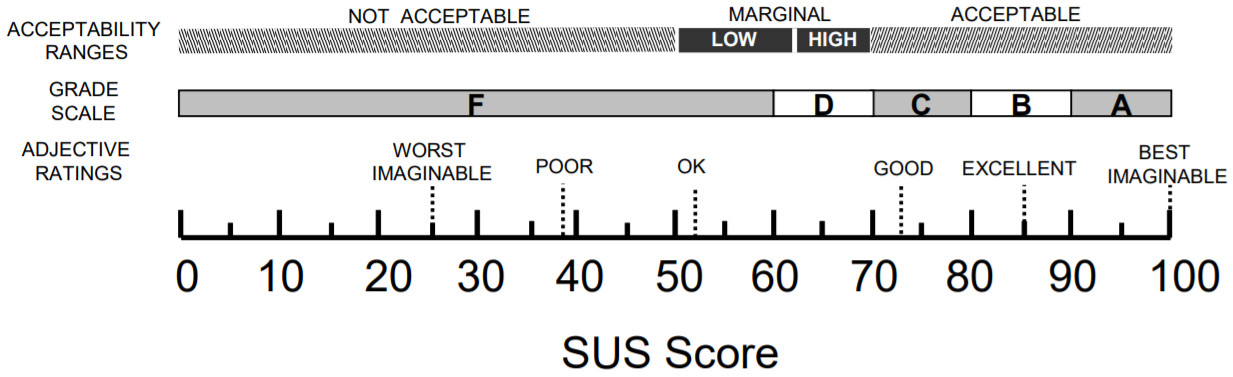}
\caption{Adjective ratings, acceptability scores, and school grading scales, in relation to the average SUS score.}
\label{fig:sus-scale}
\end{figure}

Ultimately, SUS scores have a range of 0 to 100. The SUS scale is quite intuitive. Yet, it raises many questions about its meaning in an absolute sense. To overcome this limitation, Bangor \textit{et al.} \cite{bangor2009determining} assigned grades as a function of SUS scores ranging from \textbf{F} (\textit{worst imaginable}) to \textbf{A} (\textit{best imaginable}), as shown in Figure \ref{fig:sus-scale}.

SUS has become an indispensable component in the field of user experience (UX), shedding light on the usability and, in a broader context, the overall quality of digital platforms. Usability testing with the SUS tool has also been widely adapted for mobile applications , which have attracted considerable attention from researchers and practitioners due to the high proliferation of mobile devices \cite{weichbroth2020usability}.

In addition, the scope of SUS goes beyond the UX domain. So far, the SUS has been also adopted and adopted in other industries, including automotive \cite{trojaniello2018evaluating, selinka2020usability}, 
construction \cite{ogunseiju2022mixed, olayiwola2023design}, 
e-commerce \cite{akbar2023measuring, islam2023seer}, 
education \cite{pal2020perceived, ozgen2021usability}, 
energy \cite{klein2017need, wu2022linkclimate}, 
finance \cite{ravendran2012tag, kausar2022gra}, 
healthcare \cite{bokhari2016development, tuah2021mydiabetes}, 
manufacturing \cite{andalibi2021visualization, zhang2024unlocking}, 
technology \cite{ivanova2022map, weichbroth2024usability}, 
transportation \cite{Rustamov2020, golightly2022ride2rail}, 
just to name a few. 

Therefore, our software package can be used by researchers in a wide variety of fields, enabling them to efficiently analyze data, automate calculation process, and gain visual insights, regardless of their level of expertise. In this view, the main contribution lays in developing a open-source software for usability studies, regardless of the applied experimental setup and settings.

\section{Software description}
The main achievement of Sus-Lib is the ability to calculate the software usability scale (SUS), which is collected from the users through a questionnaire. The software package is implemented in a user-friendly and open source tool. 
It is designed to be easy to use and therefore requires no special Python knowledge to use. 

\subsection{Software architecture}
Sus-Lib is a standalone Python application. While its straightforward capabilities for processing semicolons data are particularly noteworthy, the core functionality of our software package is to use such datasets to calculate individual usability scores for the evaluated object. Full installation and usage instructions are available on the Github software repository.
\\Sus-Lib consists of 4 files:
\begin{enumerate}[label=\alph*.]
    \item \textbf{setup.py}: This file uses setuptools to configure the packaging of sus\_lib, version 0.1. It includes the sus\_lib package and specifies dependencies on matplotlib and numpy.
    \item \textbf{readme.md}:  This file contains information about the other files in a directory.
    \item \textbf{sus\_lib.py}: This file contains all the functions of the library.
    \item \textbf{\_\_init\_\_.py}: This file imports functions from the sus.py.
\end{enumerate}

Sus-Lib consists of the following 11 functions:

\begin{enumerate}[label=\alph*.]
    \item \textbf{import\_from\_csv(answers=None)}
    \begin{description}
        \item[\textbf{Input:}]
        \texttt{answers}: Optional parameter. If provided, it should be either None or a string representing the path to a CSV file. In other cases, the input remains unchanged and is returned as the result.
        
        \item[\textbf{Output:}] 
        Returns a list of lists where each sublist represents a row from the CSV file, with each value converted to an integer.
        
        \item[\textbf{How it works:}] 
        If \texttt{answers} is not provided or is None, the function opens a file dialog for the user to select a CSV file. If a file path is provided as \texttt{answers}, it attempts to open that file. \\
        The function reads the CSV file, assuming a semicolon delimiter, and converts each value to an integer. If the file is not found, it raises a ValueError.
    \end{description}
    
    \item \textbf{calculate\_sus\_values(answers=None)}
    \begin{description}
        \item[\textbf{Input:}]
        \texttt{answers}: Optional parameter. Can be None, a string representing the path to a CSV file with data, a single list of 10 integers from 1 to 5 representing SUS responses or a list of them.
        
        \item[\textbf{Output:}] 
        Returns the SUS score as a float or a list of floats if \texttt{answers} is a list of lists.
        
        \item[\textbf{How it works:}] 
        The function calls \texttt{import\_from\_csv} to load the data if necessary. If answers is a list of lists, it recursively calculates the SUS score for each sublist.
        For each list of 10 integers, it calculates the SUS score using the formula:
        \begin{equation}
           SUS = 2.5 * (X + Y)
        \end{equation}
        where:
        
            \begin{equation}
                X = \sum (i_1 - 1, i_3 - 1, i_5 - 1, i_7 - 1, i_9 - 1)
            \end{equation}
                    \begin{equation}
                Y = \sum (5 - i_2, 5 - i_4, 5 - i_6, 5 - i_8, 5 - i_{10})
            \end{equation}

    \end{description}

    \item \textbf{show\_statistics(answers=None, output\_path: str="results.txt")}
    \begin{description}
        \item[\textbf{Input:}]
        \texttt{answers}: Optional parameter. Can be None, a string representing the path to a CSV file with data, a single list of 10 integers from 1 to 5 representing SUS responses or a list of them.\\ 
        \texttt{output\_path}: Optional parameter. Can be a string.
        
        \item[\textbf{Output:}] 
        Displays statistics to the console and saves them to a specified text file.
        
        \item[\textbf{How it works:}] 
        The function calls \texttt{import\_from\_csv} to load the data if necessary. Afterwards it calculates SUS values using the \texttt{calculate\_sus\_values} function. If the result is a single floating-point number (indicating only one SUS value), it calculates the following statistics: SUS value, acceptability, grade and adjective. In case multiple SUS values are present, it calculates additional statistics: mean, standard deviation, quartiles, acceptability ranges, grades, and adjective ratings using numpy and other sus-library functions such as \texttt{calculate\_sus\_values}, \texttt{calculate\_acceptabilities}, \texttt{calculate\_grades} or \texttt{calculate\_adjectives}. The statistics are formatted as a string and printed to the console. The function checks if output\_path is a valid string. If not, it opens a file dialog for the user to select where to save the file. If an error occurs during file saving, an exception message is printed.

    \end{description}

    \item \textbf{sus\_value\_histogram(answers=None)}
    \begin{description}
        \item[\textbf{Input:}]
        \texttt{answers}: Optional parameter. Can be None, a string representing the path to a CSV file with data, a single list of 10 integers from 1 to 5 representing SUS responses or a list of them.
        
        \item[\textbf{Output:}]
        Displays a histogram of SUS values.
        
        \item[\textbf{How it works:}] 
        The function first loads the data if necessary using \texttt{import\_from\_csv}. It calculates the SUS values and then generates a histogram using matplotlib with 10 intervals ranging from 0 to 100.

    \end{description}

    \item \textbf{acceptability\_bar\_chart(answers=None)}
    \begin{description}
        \item[\textbf{Input:}]
        \texttt{answers}: Optional parameter. Can be None, a string representing the path to a CSV file with data, a single list of 10 integers from 1 to 5 representing SUS responses or a list of them.
        
        \item[\textbf{Output:}] 
        Displays a bar chart of the acceptability ranges.
        
        \item[\textbf{How it works:}] 
        The function first loads the data if necessary using \texttt{import\_from\_csv}. It calls \texttt{calculate\_acceptability} to calculate the SUS values and determine the corresponding acceptability range for each. A bar chart is created showing the frequency of each acceptability range using matplotlib.
       
    \end{description}

    \item \textbf{grade\_bar\_chart(answers=None)}
    \begin{description}
        \item[\textbf{Input:}]
        \texttt{answers}: Optional parameter. Can be None, a string representing the path to a CSV file with data, a single list of 10 integers from 1 to 5 representing SUS responses or a list of them.
        
        \item[\textbf{Output:}] 
        Displays a bar chart of the SUS grades.
        
        \item[\textbf{How it works:}] 
        The function first loads the data if necessary using \texttt{import\_from\_csv}. It calls \texttt{calculate\_grade} to calculate the SUS values and assign a grade (A, B, C, D, F) to each based on the value. A bar chart is created showing the frequency of each grade using matplotlib.
    
    \end{description}

    \item \textbf{adjective\_bar\_chart(answers=None)}
    \begin{description}
        \item[\textbf{Input:}]
        \texttt{answers}: Optional parameter. Can be None, a string representing the path to a CSV file with data, a single list of 10 integers from 1 to 5 representing SUS responses or a list of them.
        
        \item[\textbf{Output:}] 
        Displays a bar chart of adjective ratings based on SUS values.
        
        \item[\textbf{How it works:}] 
        The function first loads the data if necessary using \texttt{import\_from\_csv}. It calls \texttt{calculate\_adjective} to calculate the SUS values and determine the corresponding adjective rating for each. A bar chart is created showing the frequency of each adjective rating using matplotlib.
    
    \end{description}

    \item \textbf{calculate\_acceptability(value: float)}
    \begin{description}
        \item[\textbf{Input:}]
        \texttt{value}: A single SUS value as a float.
        
        \item[\textbf{Output:}] 
        Returns a string representing the acceptability range (e.g., "NOT ACCEPTABLE", "ACCEPTABLE").
        
        \item[\textbf{How it works:}] 
        Based on the SUS value, the function determines the corresponding acceptability range using defined thresholds.
        \begin{table}[h!]
        \centering
        \begin{tabular}{|c|c|}
        \hline
        \textbf{Range of Values} & \textbf{Description} \\ \hline
        \texttt{SUS value < 50} & NOT ACCEPTABLE \\ \hline
        \texttt{50 $\leq$ SUS value < 62.5} & LOW MARGINAL \\ \hline
        \texttt{62.5 $\leq$ SUS value < 70} & HIGH MARGINAL \\ \hline
        \texttt{SUS value $\geq$ 70} & ACCEPTABLE \\ \hline
        \end{tabular}
        \caption{Classification of values based on acceptability}
        \label{table:acceptability}
        \end{table}
    
    \end{description}

    \item \textbf{calculate\_acceptabilities(answers=None)}
    \begin{description}
        \item[\textbf{Input:}]
        \texttt{answers}: Optional parameter. Can be None, a string representing the path to a CSV file with data, a single list of 10 integers from 1 to 5 representing SUS responses or a list of them.
        
        \item[\textbf{Output:}] 
        Returns a list of acceptability ranges corresponding to the SUS values.
        
        \item[\textbf{How it works:}] 
        The function calculates the SUS values and then maps each to its acceptability range using \texttt{calculate\_acceptability}.
    
    \end{description}

    \item \textbf{calculate\_grade(value: float)}
    \begin{description}
        \item[\textbf{Input:}]
        \texttt{value}: A single SUS value as a float.
        
        \item[\textbf{Output:}] 
        Returns a string representing the grade ("A", "B", "C", "D", "F").
        
        \item[\textbf{How it works:}] 
        Based on the SUS value, the function determines the grade using defined thresholds.
        \begin{table}[h!]
        \centering
        \begin{tabular}{|c|c|}
        \hline
        \textbf{Range of Values} & \textbf{Description} \\ \hline
        \texttt{SUS value < 60} & F \\ \hline
        \texttt{60 $\leq$ SUS value < 70} & D \\ \hline
        \texttt{70 $\leq$ SUS value < 80} & C \\ \hline
        \texttt{80 $\leq$ SUS value < 90} & B \\ \hline
        \texttt{SUS value $\geq$ 90} & A \\ \hline
        \end{tabular}
        \caption{Classification of values based on grade}
        \label{table:grade}
        \end{table}

    \item \textbf{calculate\_grades(answers=None)}
    \begin{description}
        \item[\textbf{Input:}]
        \texttt{answers}: Optional parameter. Can be None, a string representing the path to a CSV file with data, a single list of 10 integers from 1 to 5 representing SUS responses or a list of them.
        
        \item[\textbf{Output:}] 
        Returns a list of grades corresponding to the SUS values.
        
        \item[\textbf{How it works:}] 
        The function calculates the SUS values and then maps each to a grade using \texttt{calculate\_grade}.
    
    \end{description}
    
    \end{description}
    \item \textbf{calculate\_adjective(value: float)}
    \begin{description}
        \item[\textbf{Input:}]
        \texttt{value}: A single SUS value as a float.
        
        \item[\textbf{Output:}] 
        Returns a string representing the adjective rating (e.g., "WORST IMAGINABLE", "BEST IMAGINABLE").
        
        \begin{table}[h!]
        \centering
        \begin{tabular}{|c|c|}
        \hline
        \textbf{Range of Values} & \textbf{Description} \\ \hline
        \texttt{SUS value < 25} & WORST IMAGINABLE \\ \hline
        \texttt{25 $\leq$ SUS value < 39} & POOR \\ \hline
        \texttt{39 $\leq$ SUS value < 52} & OK \\ \hline
        \texttt{52 $\leq$ SUS value < 73} & GOOD \\ \hline
        \texttt{73 $\leq$ SUS value < 85} & EXCELLENT \\ \hline
        \texttt{SUS value $\geq$ 85} & BEST IMAGINABLE \\ \hline
        \end{tabular}
        \caption{Classification of values based on acceptability}
        \label{table:adjective}
        \end{table}
    
    \end{description}

    \item \textbf{calculate\_adjectives(answers=None)}
    \begin{description}
        \item[\textbf{Input:}]
        \texttt{answers}: Optional parameter. Can be None, a string representing the path to a CSV file with data, a single list of 10 integers from 1 to 5 representing SUS responses or a list of them.
        
        \item[\textbf{Output:}] 
        Returns a list of adjective ratings corresponding to the SUS values.
        
        \item[\textbf{How it works:}] 
        The function calculates the SUS values and then maps each to an adjective rating using \texttt{calculate\_adjective}.

    \end{description}  
    
\end{enumerate}

 \subsection{Software functionalities}
\begin{enumerate}[label=\alph*.]
\item \textbf{Data importation}. Users can import SUS response data from CSV files using the import\_from\_csv function. This ensures compatibility with common data formats and ease of use.

\item \textbf{SUS score calculation}. The calculate\_sus\_values function processes individual responses (each containing 10 items rated 1 to 5) by computing the SUS score using the standard formula. 

\item \textbf{Categorization}. Functions like calculate\_acceptabilities, calculate\_grades, and calculate\_adjectives categorize SUS scores into groups, providing quick insights into the overall usability of the product.

\item \textbf{Statistics}. The show\_statistics function generates a detailed statistical report, including mean, median, standard deviation, and quartiles. It also breaks down the data into acceptability ranges, grades, and adjective ratings.

\item \textbf{Data visualization}. The library includes functions such as sus\_value\_histogram, acceptability\_bar\_chart, grade\_bar\_chart, and adjective\_bar\_chart, allowing users to create visual representations of the data. These charts help in identifying patterns and trends within the SUS scores.

\end{enumerate}
  

\section{Illustrative example}
Let us now consider a hypothetical case study that involves a usability evaluation of an abstract object. 
In this case, a researcher has collected twenty responses using the SUS tool, administered to twenty individual participants. In this scope, Listing \ref{data:label} shows a detailed view of a used input.
\lstset{basicstyle=\footnotesize\ttfamily,language={}}

\lstset{
    caption={Collected twenty responses to a 10-item SUS questionnaire},
    label={data:label}
}
\begin{lstlisting}
5;2;5;1;4;1;5;1;4;2
5;2;5;1;4;1;5;1;4;1
5;2;5;1;4;1;5;1;5;5
5;2;5;1;4;1;5;1;4;2
5;1;4;1;5;2;4;1;4;4
4;1;5;2;5;1;5;2;5;1
5;2;5;1;4;1;5;1;4;2
5;2;5;1;4;1;5;1;4;1
5;2;5;1;4;1;5;1;5;5
5;2;5;1;4;1;5;1;4;2
5;1;4;1;5;2;4;1;4;4
4;1;5;2;5;1;5;2;5;1
5;1;4;1;5;2;4;1;4;4
1;4;2;5;1;4;3;3;2;4
4;1;5;2;5;1;5;2;5;1
5;2;5;1;4;1;5;1;4;2
5;2;5;1;4;1;5;1;4;1
4;1;5;2;5;2;4;2;5;1
2;4;5;4;3;3;1;4;2;2
1;5;1;5;2;4;1;5;1;5
\end{lstlisting}

In a typical scenario, after defining an input, a user first requests that SUS scores be calculated. Second, basic descriptive statistics are estimated for the calculated SUS scores, along with the 
frequency distribution for acceptability ranges, grade scale, and adjective ratings. 
For a given input, the numerical report, describing the usability evaluation, first presents the mean, standard deviation, first quartile (Q1), median (Q2), and third quartile (Q3), followed by three frequency distributions. Finally, the summary is displayed, which consists of a list showing each SUS score with the assigned acceptability range, grade scale, and adjective rating. 

In this regard, Listing \ref{data:summary} presents a numerical report of the usability evaluation study.

\lstset{
    caption={The numerical report for the usability evaluation},
    label={data:summary}
}
\begin{lstlisting}
SUS values
-----------
90.0
92.5
85.0
90.0
82.5
92.5
90.0
92.5
85.0
90.0
82.5
92.5
82.5
22.5
92.5
90.0
92.5
87.5
40.0
5.0


Statistic           Value               
--------------------------
Mean                78.88               
Standard Deviation  25.20               
First Quartile (Q1) 82.50               
Median (Q2)         90.00               
Third Quartile (Q3) 92.50               


Acceptability       Number              
---------------------------
NOT ACCEPTABLE      3                   
LOW MARGINAL        0                   
HIGH MARGINAL       0                   
ACCEPTABLE          17                  


Grades              Number              
--------------------------
A                   11                  
B                   6                   
C                   0                   
D                   0                   
F                   3                   


Adjectives          Number              
--------------------------
WORST IMAGINABLE    2                   
POOR                0                   
OK                  1                   
GOOD                0                   
EXCELLENT           3                   
BEST IMAGINABLE     14                  


SUS Value      Acceptability  Grade          Adjective      
------------------------------------------------------------
90.00          ACCEPTABLE     A              BEST IMAGINABLE
92.50          ACCEPTABLE     A              BEST IMAGINABLE
85.00          ACCEPTABLE     B              BEST IMAGINABLE
90.00          ACCEPTABLE     A              BEST IMAGINABLE
82.50          ACCEPTABLE     B              EXCELLENT      
92.50          ACCEPTABLE     A              BEST IMAGINABLE
90.00          ACCEPTABLE     A              BEST IMAGINABLE
92.50          ACCEPTABLE     A              BEST IMAGINABLE
85.00          ACCEPTABLE     B              BEST IMAGINABLE
90.00          ACCEPTABLE     A              BEST IMAGINABLE
82.50          ACCEPTABLE     B              EXCELLENT      
92.50          ACCEPTABLE     A              BEST IMAGINABLE
82.50          ACCEPTABLE     B              EXCELLENT      
22.50          NOT ACCEPTABLE F              WORST IMAGINABLE
92.50          ACCEPTABLE     A              BEST IMAGINABLE
90.00          ACCEPTABLE     A              BEST IMAGINABLE
92.50          ACCEPTABLE     A              BEST IMAGINABLE
87.50          ACCEPTABLE     B              BEST IMAGINABLE
40.00          NOT ACCEPTABLE F              OK             
5.00           NOT ACCEPTABLE F              WORST IMAGINABLE

\end{lstlisting}

\begin{figure}[!htb]
    \centering
    \includegraphics[width=0.8\linewidth]{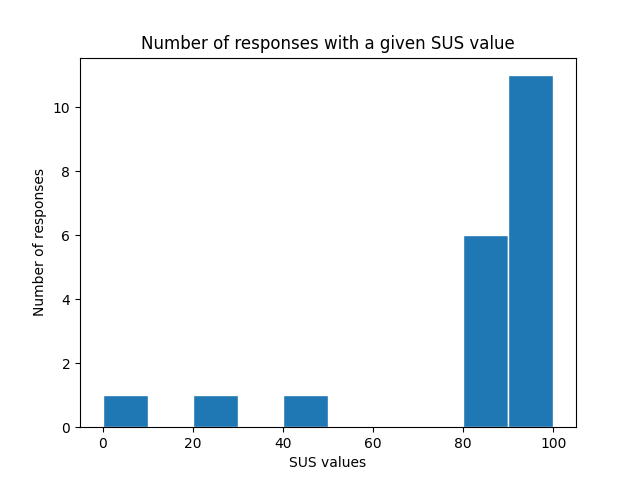}
    \caption{SUS value histogram}
    \label{fig:SUS-histogram}
\end{figure}

\begin{figure}[!htb]
    \centering
    \includegraphics[width=0.8\linewidth]{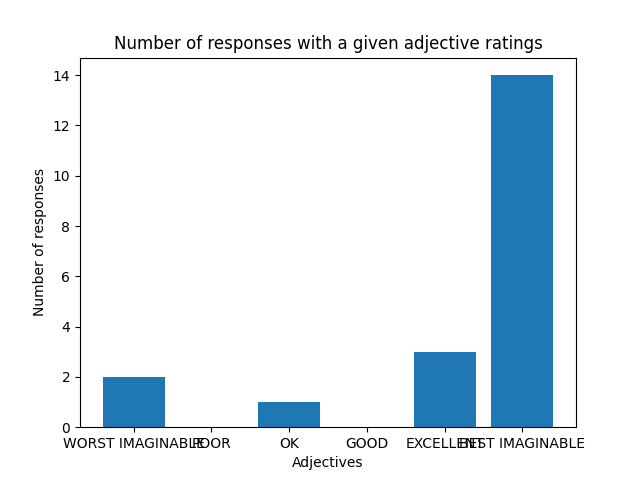}
    \caption{Adjective ratings chart}
    \label{fig:Adjective-chart}
\end{figure}

\begin{figure}[!htb]
    \centering
    \includegraphics[width=0.8\linewidth]{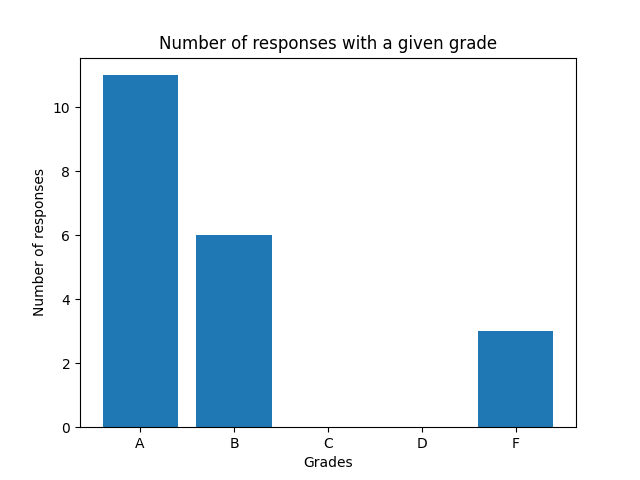}
    \caption{Grade chart}
    \label{fig:Grade-chart}
\end{figure}

\begin{figure}[!htb]
    \centering
    \includegraphics[width=0.8\linewidth]{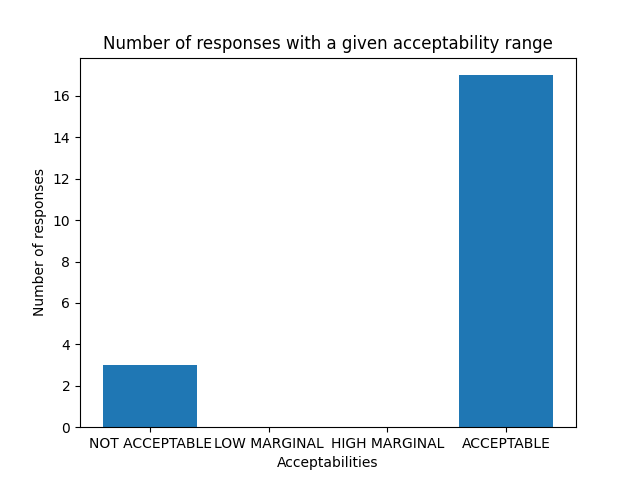}
    \caption{Acceptability level chart}
    \label{fig:acceptability-chart}
\end{figure}

Finally, with the goal of curating the data into a form that is easier to understand, highlighting the patterns, trends, and outliers, a user wishes to visualize the results in terms of the frequency distribution of the calculated SUS scores (Figure~\ref{fig:SUS-histogram}), along with three comparisons of categorical variables, namely: adjective ratings (Figure~\ref{fig:Adjective-chart}), grades (Figure~\ref{fig:Grade-chart}), and acceptability levels (Figure~\ref{fig:acceptability-chart}).


\section{Impact}
Usability evaluation tools are essential for carrying out empirical studies applicable to real information systems, regardless of architecture, complexity, or platform. While each tool has strengths and weaknesses in specific configurations and settings, the generic nature of the SUS tool has made it the most applicable tool to date. 

Sus-Lib makes calculations and visualizations fast and easy to perform by providing a single interface that includes a complete list of functions, regardless of the input size. In other words, it turns a previously tedious spreadsheet-based analysis task into a straightforward command-line sequence of requests, semantically consistent with the provided research context.

From a user perspective, by simplifying the course of action, along with the necessary workload, Sus-Lib not only eases the efforts of computer science researchers, but also makes the SUS tool more accessible to a broader audience by requiring only a basic level of command-line skills. Therefore, the Sus-Lib responds to the demand for intuitive, user-oriented and open source software by taking advantage of the Python programming language, known for its clear and easy-to-understand syntax.

The impact of Sus-Lib extends to researchers and practitioners in the field of human-computer interaction. While considering the former group, designing, exploring, and ultimately testing new solutions is at the heart of research and development efforts \cite{baumgartner2021questionnaire}, as looking at the former group, adopting, implementing, and deploying in the end-user solutions made the software business profitable and growing. In this sense, both quantitative and visual outcome, provided by the Sus-Lib software, are necessary to make informed decisions related to usability evaluation \cite{hertzum2012usability, weichbroth2022empirical, maqbool2023potential}.

\section{Conclusions}
This paper has discussed the Sus-Lib software, intended to perform all necessary calculations in usability evaluation studies, along with visualization capabilities. Due to the use of open source technologies, Sus-Lib is a cost-effective solution, requiring minimal knowledge and effort from the user on the one hand, and considerable low hardware resources on the other. In addition, the availability of the source code makes it possible to verify the implementation of the usability evaluation method, providing the sense of using reliable and trustworthy software. In this way, Sus-Lib contributes to the ongoing scientific effort to develop and share sustainable, open and reproducible methods and tools.

\section*{Declaration of competing interest}
The authors declare that they have no known competing financial interests or personal relationships that could have appeared to influence the work reported in this paper.

\bibliographystyle{elsarticle-num}

\end{document}